\documentclass[review]{elsarticle}

\usepackage{lineno,hyperref}
\modulolinenumbers[5]

\journal{Journal of Atmospheric and Solar-Terrestrial Physics}

\begin{document}

\begin{frontmatter}

\title{Relating photometric and magnetic properties of structures at solar surface}
%\tnotetext[mytitlenote]{Fully documented templates are available in the elsarticle package on &\href{http://www.ctan.org/tex-archive/macros/latex/contrib/elsarticle}{CTAN}.}

\author{Vladimir Obridko}
\address{IZMIRAN, 4, Kaluzhskoe Shosse, Troitsk, Moscow 108840, Russia}
\author{Dmitry Sokoloff}
\address{Department of Physics, Moscow State University, 119991, Russia and IZMIRAN, 4, Kaluzhskoe Shosse, Troitsk, Moscow 108840, Russia }
\ead{sokoloff.dd@gmail.com}
\author{Maria Katsova}
\address{Sternberg State Astronomical Institute, Lomonosov Moscow State University, 13, Universitetsky Prospekt, Moscow 119991, Russia}

%\author[mymainaddress,mysecondaryaddress]{Dmitry Sokoloff}

%\author[mysecondaryaddress]{Dmitry Sokoloff}

\begin{abstract}
We investigate sharp structures visible in solar magnetic field tracers.
It is shown that the sunspot magnetic boundaries do not coincide with the photometric ones. Moreover, there is no clear boundary of the magnetic field in the vicinity of sunspots. Thus, the widely accepted concept of magnetic tubes with sharp edges is not always correct and should be used with caution. It is also shown that even in the moments of complete absence of visible spots on the Sun, there are magnetic fields over 800 Gauss. The nature of these strong magnetic fields remains unclear; they may originate at relatively small depths under the photosphere. 
%Поэтому наличие резких фотометрических границ у структур на Солнце и в других небесных телах, где может происходить  образование волокон,  не обязательно указывает на наличие таких же резких магнитных границ.
\end{abstract}


\end{frontmatter}

%\linenumbers
\section{Introduction}

From the very beginning and long afterwards, the number and area of sunspots were determined visually from solar images based on their photometric properties. Nowadays, we are using photo and numerical records. However, in all cases, the data on a sunspot area are based on the image of the sunspot and the photometric estimate of its boundary (see e.g. \cite{BL64, BandS69, Jetal19}).
	There is no doubt that the main factor determining the very existence of a sunspot is the magnetic field. Nevertheless, a definition of the sunspot boundary in terms of the magnetic field is still not sufficiently elaborated in scientific literature. In our recent paper \cite{Oetal22}, we considered a related problem and our intention here is to go further in this direction.

A remarkable fact here is that most of the objects at the solar surface have a sharply defined photometric boundary. The point is that the horizontal optical thickness is quite short (about 100 km) at least in the photosphere, and the horizontal optical transport is rather difficult.  Based on the magnetic nature of almost all surface solar objects, the concept of sharp magnetic boundaries is widely anticipated.  

In particular, for quite a long time, the magnetic field outside sunspots was considered negligible. So, equations were derived, according to which the magnetic field vanishes at the outer boundary of the penumbra \cite{B42, M53}, and the dependence of the field intensity on the distance from the center of a symmetric spot was fully determined by the maximum magnetic field at the center. Later, various estimates for $B_b/B_0 = c$ were adopted (here $B_0$ is the magnetic field at the sunspot center and $B_b$ is the field at the sunspot boundary), in particular, $c = 0.5$ \cite{BS69}, $c = 0.2$ \cite{W74}, 
$c = 0.163$ \cite{GH81}, and $c = 0.607$ \cite{K83}.

The assumption that the magnetic and photometric boundaries coincide, which still needs verification, resulted, nevertheless, in a theoretical concept of magnetic tubes and ropes. 

Nowadays, the concept of a floating magnetic tube is widely accepted. It is believed, that sunspots arise during the formation of active regions on the solar surface from a strong toroidal field generated by the solar dynamo. In fact, all arguments in favor of this concept are based on theoretical considerations \cite{Cetal95, Cetal98, FF14, F08, Wetal11, Getal16, GB19}. A critical analysis of the mechanism of formation of sunspots and, more broadly, bipolar ARs  described  above has been recently performed in \cite{K09, K13, Setal13, SK14, RC14, Getal16, GB19, Zetal22}. 

The concept of magnetic ropes in the solar corona based on observations of the filament structure similar to the structure of magnetic lines is also widely accepted. It is, however, difficult to prove that this structure indeed consists of isolated tubes, because there are no direct magnetic field observations therein. The observed photometric feature may be associated with moderate variations in the magnetic field  while the large-scale magnetic field on the whole remains quasi homogeneous. The magnetic-field variations can substantially affect the coronal plasma radiation due to a strong dependence of radiation mechanisms on the  field intensity  (e.g. \cite{P82}).

\section{The sunspot magnetic boundary given by observational data}

 We proposed \cite{Oetal22} a new method for obtaining the magnetic boundary of visible sunspots based on long-term
 data series . We used SDO/HMI data on the daily longitudinal magnetic field for 2375
 days from 01.05.2010 to 31.10.2016. 
 We use the daily sunspot numbers from \href{http://sidc.oma.be/silso/datafiles}{WDC–SILSO, Royal Observatory of Belgium, Brussels (version 2)}. 
 The cumulative daily sunspot areas were taken from the \href{https://solarscience.msfc.nasa.gov/greenwch.shtml}{NASA Web site}. 
 At present, there are two databases formed of high–resolution observations carried out with single–type instruments. These are SOHO/MDI and SDO/HMI. Michelson Doppler Imager (MDI) onboard the Solar and Heliospheric Observatory (SOHO) \cite{Setal95} was continuously measuring the Doppler velocity, longitudinal magnetic field, and brightness of the Sun for 15 years up to 12 April 2011. The enhanced Helioseismic and Magnetic Imager (HMI: \cite{Setal12}) onboard the Solar Dynamics Observatory(SDO: \cite{Petal12}) started its routine observations on 30 April 2010. HMI data include all MDI observables, but with much higher spatial and temporal resolutions and better data quality. 

 We find the relative area at the solar surface occupied by the magnetic field larger than a certain threshold value. This relative area is expressed in millionths of the visible
hemisphere (m.v.h.), as is customary when studying the total sunspot areas.
These calculations are compared with the database of daily sunspot observations. Putting together all of the data, we arrive at the conclusion that, on
average, the magnetic boundary of a sunspot as defined by the normal component of the magnetic field is 550 G. This estimate is quite reliable. Indeed,  this value provides maximum for correlation between magnetic and visual data. Even rather small variations of the value chosen to determine the boundary (say, 525 G or 575 G instead of 550 G because 575 is greater than 550) reduces the correlation substantially
(\cite{Oetal22}).

The point is that the magnetic field at the apparent sunspot boundary does not vanish. Discussing the situation in terms of the magnetic tubes, we must have in mind that a magnetic tube has no sharp boundary and extends far beyond the photometric sunspot boundary. The relation between the magnetic field strength in a circular sunspot versus its relative photometric  radius normalized to the radius sharp optical spot boundary  with $\rho_0=1$ at the photometric boundary is shown in Fig.~1.

The link between the magnetic field (measured in G, Fig.~1 shows  $\log B$) and the relative photometric sunspot radius 
was calculated up to $\rho = 1.2$ and was approximated by the second order polynomial as 

\begin{equation}
\log B=3,39537-0,90097 \rho +0,25188 \rho^2, 
\label{app}
\end{equation}
The magnetic field becomes as low as several hundred G at $\rho \approx 2$, where it already does not differ from the magnetic field of surrounding spots and faculae. An exact estimate is not possible here, since  the faculae usually have a diffuse non-axisymmetric form; so it is better to speak about a complex of a sunspot and surrounding faculae rather than a photometric spot. That is why the approximation in Fig.~1 is extrapolated up to $\rho = 1.8$ using Eq.~(1) and shown by dashed line. Perhaps this extrapolation slightly overestimates the magnetic field, and at the faculae boundary, it is as low as several dozens of gauss. 

\begin{figure}[p]
{\includegraphics[width=\textwidth]{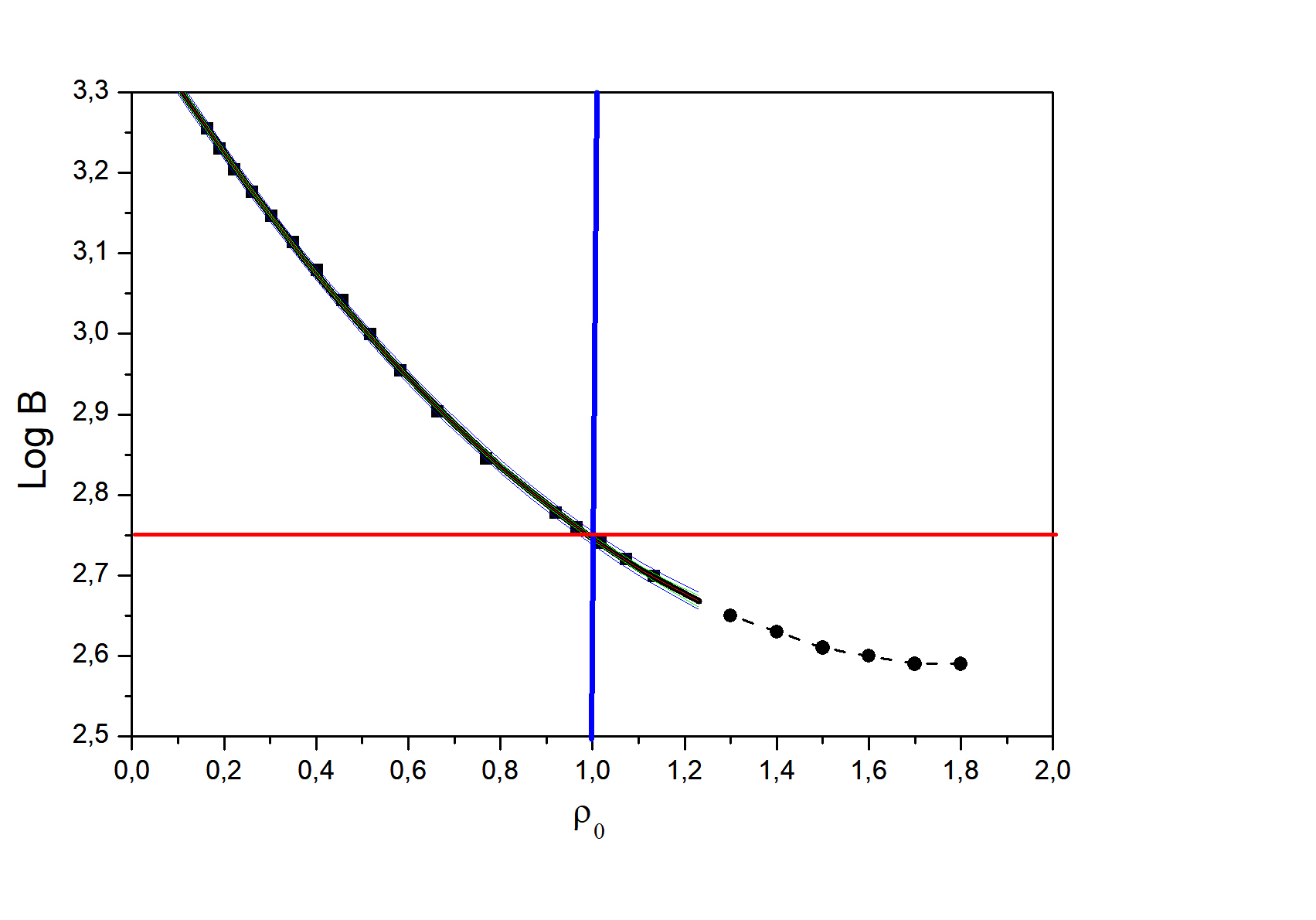}}
\caption{Empirical dependence of the magnetic field (dots, measured in G) in a symmetric sunspot on the relative sunspot radius ($\rho =1$ at the photometric boundary). The solid line is a polinomial approximation in Eq.~(1), the dashes extrapolate the approximation for weak magnetic fields. The thin blue lines near the approximation mark the 95\% confidence interval. The thick blue line shows the photometric  sunspot boundary, and the red line shows the magnetic sunspot radius, calculated presuming that the magnetic radius of the sunspot corresponds to the magnetic field strength of 550 G.}
\label{fig:f1}
\end{figure}

\begin{figure}[p]
{\includegraphics[width=\textwidth]{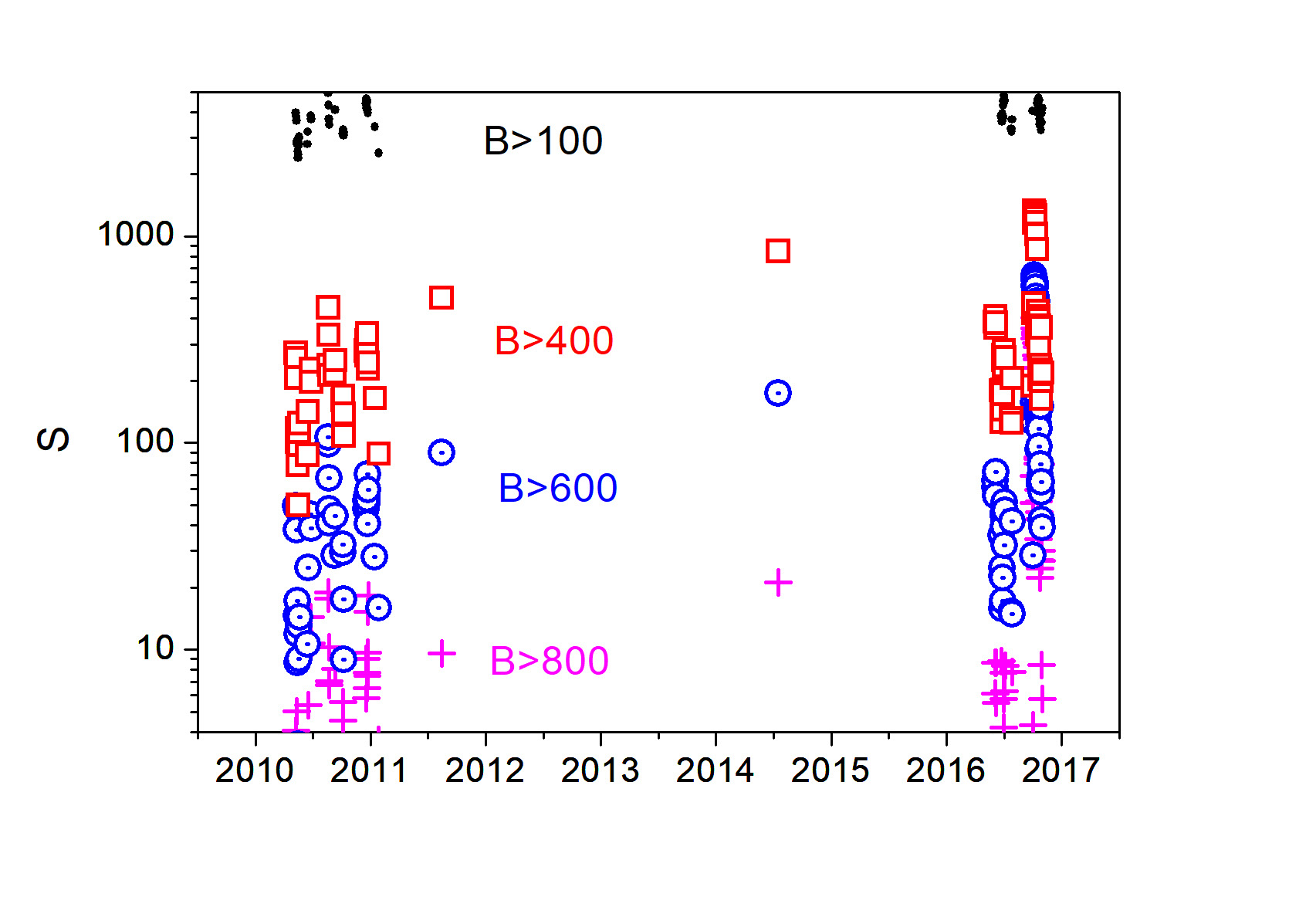}}
\caption {Statistics of the magnetic-field strengths on spotless days. The black dots stand for $B>100\;$G, the red squares stand for $B>400\;$G,  the blue dotted circles give data for $B>600\;$G, and  purple crosses indicates the area covered by strong magnetic field $B>800\;$G.}
\label{fig:f2}
\end{figure}

%%%%%%%%%%%%%%%%%%%%%%%%%%%%%%%%%%%%%%%%%%%%%%%%%%%%%%%%

We conclude that the apparent sunspot boundary is determined by interaction of the magnetic field and the convective transport. If the magnetic field becomes low enough to suppress convection in all scales however sufficient to suppress convection in small scales, the  convective transport becomes slower and the brightness of the element decreases \cite{P60}\footnote{Here, we give reference (correcting a misprint) to this quite old, but helpful paper as it is given in ADS. In fact, however, the paper was translated to English as Sov. Astron. {\bf 4}, 59, 1961 and this English translation is provided by ADS.}. The idea of the paper can be shortly described as follows. In accordance with \cite{S59}, the author claims that while a weak magnetic field does not affect the main streams, it does reduce the turbulence so that the flow becomes more stable. As a result, the convective velocity is determined by the balance between the upward force and the force of turbulent viscosity. This, in turn, decreases the dissipation and increases the convective velocities. In the process, the brightness of the plage surrounding the sunspot somewhat increases. A further increase of the magnetic field intensity leads to suppression of the convective motions. The balance condition provides the apparent sharp boundary under discussion. 

\section{Locally strong magnetic fields and magnetic field boundary}

The discussion above is aimed at  determining the sunspot boundary as a whole. There is, however, a local aspect of the problem. The point is that interaction between the magnetic field and the convective transport depends on the size of magnetic element. If the element is small enough (size about 100 km), the horizontal optical thickness for radiative transport becomes comparable with the geometric size of the element, i.e. magnetic tubes or ropes. Then, the horizontal transport can smooth in the scale of hundred kilometers the temperature profile, and it becomes problematic to isolate the element by its dropping brightness. That is why there are some elements with a strong magnetic field on the solar surface, even if there are no sunspots at the instant. This fact has to be taken into account when discussing the magnetic boundary of sunspots. 

The existence of small optical non-observable magnetic field elements was emphasized in \cite{S73, S82}. \cite{Shetal12} directly observed such elements in the solar polar region with extra-atmospheric high-resolution instruments.

In order to estimate the role of locally strong magnetic fields in the context of sunspot studies, we illustrate observations of elements with locally strong magnetic field obtained on spotless days (Fig. 2). The selection procedure is described above. We have calculated the area covered with the field above the given threshold value. The black dots stand for the magnetic field $B> 100\;$G; i.e. these are the regions totally unrelated to sunspots. The area of these regions is of the order of several thousand m.v.h. The areas of the regions covered with magnetic fields $B>400\;$G (squares) and $B> 600\;$G (circles) are slightly smaller, but  substantial. It is most impressive, however, that, even on spotless days, there are quite a lot, several dozens,  of objects with $B > 800$ G (crosses). Such objects should be considered as photometric sunspots; however, they are not observed by standard methods. The area of such regions is very small (several dozen m.v.h.), as well as their contribution to total magnetic flux. Still the total magnetic field energy in these regions may reach $10^{30}$ erg and they may be responsible for moderate solar flares. We want to emphasize that several dozens of such objects exist even in the epochs of very deep solar minimum. As they are not recorded by the sunspot patrol service, their size in cross section is apparently smaller than 2-3 arcsec.

Note that spotless days on the Sun are not  rare, especially in the epochs of the solar minimum. There can be several dozens and even hundreds of spotless days during a cycle; e.g. there were 311 spotless days during one year in 1913 and more than a thousand spotless days during the whole Cycle 14. Our results demonstrate that the absence of sunspots does not mean the absence the elements with strong magnetic fields. Fig.~2 shows that sometimes, the cumulative area  of strong magnetic fields on spotless days can be as large as a few dozen m.v.h. This corresponds to a sunspot of moderate size, which has to be well observable. The fact that sunspots were not recorded on those  days means that the magnetic fields existed in the form of isolated small magnetic elements, which are optically not distinguishable, but are accessible to magnetographic observations. 
They are observed by spectropolarimeter SP \cite{Letal01} of the SOT (Solar Optical Telescope, \cite{Tetal08b, Setal08, Shetal08, Ietal08}) aboard the Hinode satellite \cite{Ketal07}. 
Diffraction limited, high-polarization-sensitivity observations, which reveal the fine structure of photospheric
vector magnetic fields are performed in \cite{Setal12}. The first Hinode SOT observations of the polar areas revealed the existence of
many patchy magnetic concentrations with intrinsic field strengths of over 1 kG distributed across the
entire polar region \cite{Tetal08a}. 
The spatial resolution of instrument is about 0.32 arcsec (0.16 arcsec pixel size),
which corresponds to 200 km on the solar surface. According to \cite{Setal12}, the unipolar appearance and
disappearance suggests that the large patches are formed from and disintegrated into patches with
magnetic flux below the detection limit of the instrument. The component seen in the magnetic fluxes range $10^{15} \div  10^{16}$ Mx cm$^{-2}$
may be the tip of the iceberg of these unseen fluxes, and the large concentrations may have
formed from the inventory of small concentrations.

The role of small magnetic elements has to be somehow included in the scenario of the solar cycle. The conventional scenario is as follows. The differential rotation produces the large-scale toroidal magnetic field. The second dynamo driver restores (e.g., in the framework of the Babkock-Leighton mechanism)  the poloidal magnetic field of the opposite
sign. The formation of small magnetic elements is not inevitably included in this scheme and may be driven by turbulent processes. During the solar minima, i.e. the times when the large-scale dynamo action is weak, turbulent mechanisms may produce small magnetic elements all over the solar surface rather than near the solar equator only (see \cite{Setal12}). Note that at the reversal of the large-scale polar magnetic field,  the number of small magnetic elements with strong magnetic fields of both polarities is more or less equal as it should be in the small-scale dynamo action.

\section{Conclusion}

We performed a comparison between photometric and magnetic concerning the sunspot boundary to learn that
the sunspot is not an isolated magnetic tube. Just a sharp change of brightness occurs in the fields of about 550 G, and we see the penumbra.

	Fields with intensities more than 800 G exist regardless of whether or not  there are sunspots. However, when they are combined, the field intensifies and, starting from 550 G, the heat transfer is disturbed and the brightness decreases. 
The results obtained are of great importance for understanding the nature of magnetic field generation on the Sun and the emergence of active regions. The generally accepted notion of the magnetic field tubes is not entirely correct. It is believed that sunspots as particular entities arise during the formation of ARs on the solar surface from a strong toroidal field, which is generated by the solar dynamo mechanism at the base of the convection zone and is carried out into the photosphere. In fact, all arguments in favor of this concept are based on theoretical considerations.

The emergence of a single magnetic tube as a source of sunspots contradicts the observed field structure of a single sunspot. During the generation process, the turbulent dynamo creates many elements with different strengths. Their energy distribution changes with the phase of the cycle. But these elements are not tubes with isolated boundaries. The field in them decreases gradually with distance from the center of the element to its periphery. The photometric sharp boundaries of the spots are a the result of  influence of the magnetic field on the processes of energy transfer. The fields above 550 G strongly reduce the flow of energy from below and a sharp boundary appears \cite{P60}  (see also \cite{Ketal74}).

Moreover, spots emerge in a pre-existing magnetic environment and are includes in active regions.
Sunspot formation  is far to be a surface phenomenon only rather develops in leptocline what obviously requires further investigation and modeling (see e.g. \cite{Ketal23}).

Note one more time that our analysis being performed in surface solar structure context contains a more general message. Presence of sharp photometric boundaries for solar surface structures as well in cosmos as well do not immediately imply presence of corresponding sharp magnetic structures. 
Another point to be mentioned is that relation between radiative and magnetic properties of magnetic structures isolated in solar MHD numerical simulations obviously requires a more deep analyses.

\section*{Acknowledgments}
VNO, MMK and DDS acknowledge the support of the Ministry of Science and Higher Education of the Russian Federation under the grant 075-15-2020-780 (VNO and MMK) and 075-15-2022-284 (DDS). DDS thanks support by BASIS fund number 21-1-1-4-1.

\bibliography{main}

\end{document}